\documentclass[a4paper,12pt]{article}
\usepackage[usenames,dvipsnames]{pstricks}
\usepackage[utf8]{inputenc}
\pdfoutput=1
\usepackage{lmodern} 
\usepackage[T1]{fontenc}
\usepackage{multicol}
\usepackage{epsfig}
\usepackage[fleqn]{amsmath}
\usepackage{amsmath,amsfonts}
\usepackage{textcomp}
\usepackage{graphicx}
\usepackage{subcaption}
\usepackage{authblk}
\usepackage[font={small,it}]{caption}
\usepackage{epstopdf}
\usepackage[pdfencoding=auto, psdextra]{hyperref}
\usepackage{bookmark}
\usepackage{float}
\hypersetup{pdfauthor={some author},pdftitle={3+1 paper}}
\usepackage {ifpdf}
\usepackage[margin=0.9 in]{geometry}
\usepackage[english]{babel}

\usepackage[numbers,sort&compress]{natbib}

\title{\textbf{Investigating the sterile neutrino
parameters with QLC in 3 + 1 scenario}}
\author{\small{Gazal Sharma\footnote{gazzal.sharma555@gmail.com}}}

\author{\small{B. C. Chauhan\footnote{chauhan@associates.iucaa.in}}}

\affil{\textit{Department of Physics and Astronomical Science, }\\
\textit{ School of Physical and Material Sciences,}\\
\textit{ Central University of Himachal Pradesh (CUHP),}\\
\textit{Dharamshala, Kangra (HP),
India 176215}}
\date{}

\begin{document}

\maketitle

\begin{abstract}
In the scenario with four generation quarks and leptons and using a 3 + 1 neutrino model having one sterile and the three standard active neutrinos with a $4 \times 4$ unitary transformation matrix, $U_{PMNS_{4}}$, we perform a model-based analysis using the latest global data and determine bounds on the sterile neutrino parameters i.e. the neutrino mixing angles. Motivated by our previous results, where, in a quark-lepton complementarity (QLC) model we predicted the values of $\theta_{13}^{PMNS}=(9_{-2}^{+1})^{\circ}$ and $\theta_{23}^{PMNS}=(40.60_{-0.3}^{+0.1})^{\circ}$.  In the QLC model the non-trivial correlation
between $CKM_4$ and $PMNS_4$ mixing matrix is given by the correlation matrix $V_{c_{4}}$. Monte Carlo simulations are performed to estimate the texture of $V_{c4}$ followed by the calculation of $PMNS_4$ using the equation, $U_{PMNS_{4}}= (U_{CKM_{4}} . \psi_{4})^{-1}.V_{c_{4}}$, where $\psi_{4}$ is a diagonal phase matrix. The sterile neutrino mixing angles, $\theta_{14}^{PMNS}$, $\theta_{24}^{PMNS}$ and $\theta_{34}^{PMNS}$ are assumed to be freely varying between $(0-\pi/4)$  and obtained results which are consistent with the data available from various experiments, like No$\nu$A, MINOS, SuperK, Ice Cube-DeepCore. In further investigation, we
analytically obtain approximately similar ranges for various neutrino mixing parameters $\mid{ U_{\mu 4}}\mid ^2$ and $\mid{ U_{\tau 4}}\mid ^2$.
\end{abstract}

{\bf Keywords:} {Neutrino Mixing Angles, Quark-Lepton Complementarity, Sterile Neutrinos}

\section{Introduction}
After the completion of a few decades since the birth of Neutrino Physics and its experimental world, we are at a stage where we have unraveled various mysteries, including very strong evidence of neutrinos being massive and the existence of neutrino oscillation, but there are many issues that still need to be resolved. The recent results from Daya Bay, CHOOZ and other experiments \cite{1,2,3,4,5} on the relatively large value of $\theta_{13}^{PMNS}$, a clear $1^{st}$-order picture of the three flavor lepton mixing matrix have emerged \cite{Fogli,7,8}. So, according to the current experimental situation we have measured all the quark and charged lepton masses, and the value of the difference between the squares of the neutrino masses $\Delta m_{12}^2=m_2^2-m_1^2$ and $|\Delta m_{23}^2|=|m_3^2-m_2^2|$. We also know the value of the quark mixing angles and the mixing angles, $\theta_{12}^{PMNS}$, $\theta_{23}^{PMNS}$ and $\theta_{13}^{PMNS}$ in the lepton sector. The way the Cabibbo-Kobayashi-Maskawa $(U_{CKM})$ mixing matrix is there in quark sector, the phenomenon of lepton flavor mixing is described by a $3\times3$ unitary matrix called Pontecorvo-Maki-Nakagawa-Sakata $(U_{PMNS})$. Investigating global data fits of the experimental results, so far we have got a picture which suggests that the $U_{PMNS}$ matrix contains two large and a small mixing angles; i.e. the $\theta_{23}^{PMNS}$ $\approx$ $45^\circ$, the $\theta_{12}^{PMNS}$ $\approx$ $34^\circ$ and the $\theta_{13}^{PMNS}$ $\approx$ 
$9^{\circ}$.
 
These results when read along with the quark flavor mixing matrix $(U_{CKM})$, which is quite settled with three mixing angles that are small i.e. $\theta_{12}^{CKM}\approx 13^\circ$, $\theta_{23}^{CKM}\approx 2.4^\circ$ and $\theta_{13}^{CKM}\approx 0.2^\circ$, a disparity-cum-complementarity between quark and lepton mixing angles is noticed. Since the quarks and leptons are fundamental constituents of matter and also that of the Standard Model(SM), the complementarity between the two of them is seen as a consequence of some symmetry at high energy scale. This complementarity popularly named `Quark-Lepton Complementarity'(QLC) have been explored by several authors
\cite{Georgi:1979df,10,11,12,13,14}. The relation is quite appealing to do the theory and phenomenology; however, it is still an open question, what kind of symmetry could be there between these fundamental particles of two sectors. The possible consequences of the QLC have been widely investigated in the literature. In particular, a simple correspondence between the $U_{PMNS}$ and $U_{CKM}$ matrices have been proposed and used by several authors~\cite{Ferrandis:2004mq,16,17,18} and analyzed in terms of a correlation matrix $V_{c}$ \cite{81,Chauhan:2006im,Xing:2005ur,20}.

The fact that lepton flavors mix as well as oscillate leads to a new window of physics beyond the SM. Neutrino mixing may fill many voids of SM but still, there are few anomalies that could not be explained within the three flavor framework of neutrinos and points towards the existence of another flavor of neutrinos i.e.({\it sterile neutrino}) with a mass $\sim eV$ scale. Sterile neutrinos are the singlets of the Standard Model gauge symmetries that can couple to the active neutrinos via mixing only. Till now there are bounds on the active-sterile mixing, but there is no bound on the number of
sterile neutrinos and on their mass scales. The existence of sterile neutrinos is investigated at different mass scales by various experiments; LSND, MiniBooNE, MINOS, Daya Bay, IceCube etc..

The main motivation behind this work is basically testing our model in the $4^{th}$ generation scenario. After the successful results obtained in our previous papers, we have tried to extend our model and complete analysis in $3+1$ scenario.  Along with that the major motivating factor that pushed us towards the extension of our model in 3+1 scenario is that the results obtained in our previous works \cite{81,Chauhan:2006im,Xing:2005ur,20}(\textit{and references therein}) are quite consistent with the recent results from No$\nu$A \cite{19} and IceCube \cite{ic} which give us new ray of hope in favour of our model and its stability. Taking into account the precise results obtained from various experiments on  neutrino mixing angles for three generations our model and its predictions fit quite well. As we all are aware of the fact that with the pace of time we are entering into the better precision era, so considering that we have tried to extend our analysis to $3+1$ scenario with more accurate Wolfenstein parameters for $U_{CKM_{4}}$ and neutrino mixing parameters for $U_{PMNS_{4}}$ preserving unitarity up to $4^{th}$ order. Using all the parameters that are available from the global data analysis we tried to investigate the structure of the $4\times 4$ correlation matrix $V_{c_{4}}$, numerically. According to our investigations, there is a possibility of the existence and role of sterile neutrinos in the QLC, that helped us to give some constrained results for two sterile neutrino mixing angles i.e. $\theta_{24}^{PMNS}$ and $\theta_{34}$.

On the stability of the framework that we have used in order to carry out our entire analysis which is the extension of SM in $4^{th}$ generation, one might argue that the four generation scenarios are strongly disfavoured. This is true, unless a substantial modification is realized for the scalar sector. However, such an extension of the Standard Model (with massive neutrinos) is excluded by several authors eg. in references \cite{x,y}. As such, during the starting period of the discovery of Higgs particle, data was not so precise that the possibility of the $4^{th}$ generation coupling to the Standard Model Higgs doublet was not introduced in a different Beyond Standard Model scenarios. But, such options were ruled out as the LHC data develops gradually, which let many scenarios to go beyond SM(BSM). It has been noted that such $4^{th}$ generation is hidden during the single production of the Higgs Boson, while it shows up when one considers the double Higgs production i.e. $gg \rightarrow hh$ which can be considered in a different framework of a two Higgs doublet model (2HDM) \cite{x1,g,gg,y1}. This is the framework that we have taken to carry out our analysis which is well favored by the work done and published in 2018 \cite{35}. In that work, they show that the current Higgs data does not eliminate the possibility of a sequential $4^{th}$ generation that gets their masses through the same Higgs mechanism as the first three generations. 

In this paper, we have compared our data with the recent results provided by some ongoing experiments. Starting from the beginning of the sterile neutrino search by LSND and MiniBooNE anomalies we have covered a long distance till these recent experiments like No$\nu$A, SuperK, MINOS, and Ice Cube-DeepCore and many more for the search of sterile neutrinos. In this paper we do not comment upon or explain the generation of sterile neutrinos or the $4^{th}$ generation quarks instead we have used the QLC model and done some numerical analysis to obtain bounds on the values of sterile neutrino parameters using previously formulated $U_{PMNS_{4}}$ and $U_{CKM_{4}}$ \cite{Alok,24,25,26}.

The paper is organized in five sections as follows. In the next section {\bf\ref{sec:2}}, we describe in brief the theory of the QLC model and show how this model fits in the $3+1$ scenario. For the generation of $4 \times 4$ $V_{c4}$ matrix different parametrizations were taken for formulation of $U_{CKM4}$ and $U_{PMNS4}$ and all that is discussed in section {\bf\ref{sec:2}}. The investigation of correlation matrix ($V_{c4}$), using Monte Carlo simulation is done in section {\bf\ref{sec:3}}. The PMNS matrix followed by the constrained values of sterile neutrino angles is obtained using the model equation in section {\bf\ref{sec:4}} along with the results obtained using the QLC model are compared with bounds given by the global data analysis and various experiments. Finally, the conclusions are summarized in section {\bf\ref{sec:5}}. Various plots(contour and scattered plots) have been made in order to show the correlation between the $\theta_{24}$ \& $\theta_{34}$ and $\mid U_{\mu 4}\mid^{2}$ \& $\mid U_{\tau 4}\mid^{2}$ as well as for $\sin^{2}\theta_{24}$ and $\sin^{2}\theta_{34}$ against $\sin^{2}\theta_{12}$ and $\sin^{2}\theta_{23}$. We have also obtained the normal distribution function histograms for both $\theta_{24}$ and $\theta_{34}$ for different values of $m_{{t}\prime}= 400/600 GeV$.

\section{Theoretical Framework of QLC Model in 3+1 scenario} \label{sec:2}
The mixing of quarks and leptons have always been of great interest and remains a mystery in particle
physics. The search for symmetry or unification of quarks and leptons is one of the goals of particle physics, and many efforts are devoted toward this work. The bottom-up approach i.e., finding some phenomenological relations as well as
their explanations, gives some clues on this issue. In the SM the mixing of quark and lepton sectors is described by the matrices
$U_{CKM}$ and $U_{PMNS}$. When we observe a pattern of mixing angles of quarks and
leptons, and combine them with the pursuit for unification i.e. symmetry at some high energy
leads the concept of quark-lepton complementarity i.e. QLC.
Possible consequences of QLC have been widely investigated in the literature and in particular, a simple correspondence between the PMNS and CKM matrices have been proposed and analysed in terms of a correlation matrix $V_c$.
As long as quarks and leptons are inserted in the same representation of the underlying
gauge group at some higher energy scale, we need to include in our definition of $V_c$ arbitrary but non-trivial
phases between the quark and lepton matrices. Hence, we will generalize the relation
 \begin{equation} \label{eq}
  V_{c}= U_{CKM} \cdot U_{PMNS} \hspace{.5cm} \mbox{to} \hspace{.5cm} V_{c}= U_{CKM} \cdot \psi \cdot U_{PMNS}
 \end{equation}

\noindent where $V_{c}$ is the correlation matrix defined as a product of $U_{PMNS}$ and $U_{CKM}$.

When sterile neutrinos are introduced in the $3 + N_{s}$ schemes, where the $N_s$($N_{s} = 1$ for one sterile mixing) is the number of new mass eigenstates. For this case, we define $\psi_4$ in place of $\psi$ in equation {\bf \ref{eq}}. So, along with $\psi_4$ in QLC the above equation takes the form
\begin{equation}\label{2}
V_{c_{4}}=U_{CKM_{4}} \cdot \psi_{4} \cdot U_{PMNS_{4}}
\end{equation}
\noindent where the quantity $\psi_4$ is a diagonal matrix $\psi_{4}= diag(e^{\iota \psi{_i}})$ and the four phases of $\psi_i$ are set as free parameters because they are not restricted by present experimental evidences. The 3+1 active-sterile mixing scheme is a perturbation of the standard three-neutrino mixing in which the $3 \times 3$ unitary mixing matrix U is extended to a $4 \times 4$ unitary mixing matrix with $\mid U_{e4} \mid$, $ \mid U_{\mu 4}\mid$ and $ \mid U_{\tau 4}\mid$ which leads to the generation of $U_{PMNS_{4}}$ lepton mixing matrix. However, the addition of a $4^{th}$ generation to the standard model leads to a $4 \times 4$ quark mixing
matrix $U_{CKM_{4}}$, which is an extension of the Cabibbo-Kobayashi-Maskawa (CKM)
quark mixing matrix in the standard model.

\subsection{ \texorpdfstring{$CKM_{4}$}{CKM\_{4}} and \texorpdfstring{$PMNS_{4}$}{PMNS\_{4}} formulation}

In order to calculate the texture of $V_{c_{4}}$ we have used the $U_{CKM_{4}}$ and $U_{PMNS_{4}}$ taking reference from several works. Although the $4^{th}$ generation quarks are too heavy to produce in LHC, yet they may affect the low energy measurements, such as the quark ${t}\prime$ would contribute to $b \rightarrow s$ and $b \rightarrow d$ transitions, while the quark ${b}\prime$ would contribute similarly to $c \rightarrow u$ and t $\rightarrow$ c~\cite{Alok,24,25,26}.

The CKM matrix in SM is a $3 \times 3$ unitary matrix while in the $SM_{4}$ (this is the simplest extension of the SM, and retains all of its essential features: it obeys all the SM symmetries and does not introduce any new ones), the $U_{CKM_{4}}$ matrix is $4 \times 4$, matrix which can be shown as
\[
U_{CKM} =
\begin{bmatrix}
V_{ud} & V_{us} & V_{ub}\\
V_{cd} & V_{cs} & V_{cb}\\
V_{td} & V_{ts} & V_{tb}
\end{bmatrix}
,
\]

 \[
U_{CKM_{4}} =
\begin{bmatrix}
\tilde{V_{ud}} & \tilde{V_{us}} & \tilde{V_{ub}} & \tilde{V_{u{b}\prime}}\\
\tilde{V_{cd}} & \tilde{V_{cs}} & \tilde{V_{cb}} & \tilde{V_{c{b}\prime}}\\
\tilde{V_{td}} & \tilde{V_{ts}} & \tilde{V_{tb}} & \tilde{V_{t{b}\prime}}\\
\tilde{V_{{t}\prime d}} & \tilde{V_{{t}\prime s}} & \tilde{V_{{t}\prime b}} & \tilde{V_{{t}\prime{b}\prime}}
\end{bmatrix}
,
\]

\noindent where all the elements of the matrix have their usual meanings except for ${b}\prime$ and ${t}\prime$, which we have already defined above.
In the presence of the sterile neutrino $\nu_{s}$, the flavor
($\nu_{\alpha}, \alpha = e, \mu, \tau, s$) and the mass eigenstates ($\nu_{i}, i =
1, 2, 3, 4$) are connected through a $4 \times 4$ unitary mixing
matrix U, which depends on six complex parameters \cite{para1,para}.
 Such a matrix can be expressed as the product of six
complex elementary rotations, which define six real mixing
angles and six CP-violating phases. Of these phases
three are of the Majorana type and are unobservable in
oscillation processes, while the remaining three are
of the Dirac type. A particularly convenient choice of the parametrization
of the mixing matrix is
\begin{equation}
 U= \tilde{R_{24}}R_{34}\tilde{R_{14}}R_{23}\tilde{R_{13}}R_{12} P,
\end{equation}
\noindent where $R_{ij}$ and $\tilde R_{ij}$ represents a real and complex $4\times 4$ rotation
in the (i, j) plane, respectively containing the $2 \times 2$ sub matrices

\begin{eqnarray} \nonumber
R_{ij}^{2\times 2}=\begin{bmatrix}
C_{ij} & S_{ij}\\
-S_{ij} & C_{ij}
 \end{bmatrix},
\end{eqnarray}

\begin{eqnarray}
\tilde R_{ij}^{2\times 2}=\begin{bmatrix}
C_{ij} & \tilde S_{ij}\\
-\tilde S_{ij}^{*} & C_{ij}
 \end{bmatrix}
\end{eqnarray}

\noindent where $C_{ij}\equiv \cos\theta_{ij}$, $S_{ij}\equiv \sin\theta_{ij}$ and $\tilde S_{ij}\equiv S_{ij} e^{-\iota \phi_{ij}}$.

\section{Numerical Simulation and Methodology} \label{sec:3}
In the standard parametrization of $U_{CKM}$ and $U_{PMNS}$ in equation, we have inserted the observed/experimental values of the $U_{CKM}$ and $U_{PMNS}$ parameters, and obtained the probability density texture of the correlation matrix $(V_{c})$ using Monte Carlo method for $1$ billion shots for each variable. We have fully used freedom of the unknown parameters like $\psi$ and $\phi$ by varying them in the unconstrained spread $[0-2\pi]$ with flat distribution. Now, we write
\begin{equation}\label{eq:pmnsinv}
U_{PMNS_{4}}=(U_{CKM_{4}}\cdot \Psi_{4})^{-1}\cdot V_{c_{4}},
\end{equation}
this expression is the inverse of equation ({\bf\ref{2}}), which was used to estimate the texture of the correlation matrix $V_{c_{4}}$. Using equation {\bf\ref{eq:pmnsinv}} we have reverted back the results of the exercise in order to predict the unknown sector of $U_{PMNS}$. In the inverse equation the generalised correlation matrix $V_{c_4}$ thus obtained was basically used to be replaced by bimaximal(BM) and tribimaximal(TBM) matrices in the previous year by several authors \cite{81,Chauhan:2006im,Xing:2005ur,20}(\textit{and references therein}). 

As such, this model procedure using numerical method of Monte Carlo and freedom of unknown parameters is having merit as its predictive power shown in our previous works published in quality journals \cite{81,Chauhan:2006im,Xing:2005ur,20}.

The $U_{CKM_{4}}$ matrix can be described, with appropriate choices for the quark phases,
in terms of 6 real quantities and 3 phases. We have used the Dighe-Kim (DK) parameterization of the $CKM_4$ matrix \cite{Alok,24,25,26}. This matrix can be calculated in the form of an expansion in powers
of $\lambda$ such that each element is accurate up to a multiplicative factor of $[1 + {\cal O}(\lambda^3)]$.
The Dighe-Kim (DK) parametrization defines

\noindent $\tilde{V_{ud}}=1-\frac{\lambda^{2}}{2},
\tilde{V_{us}}=\lambda,
\tilde{V_{ub}}=A \lambda^{3} C e^{\iota \delta_{ub}},\\
\tilde{V_{u{b}\prime}}=p\lambda^{3}e^{-\iota\delta_{u{b}\prime}},
\tilde{V_{cd}}=-\lambda,
\tilde{V_{cs}}=1-\frac{\lambda^{2}}{2},\\
\tilde{V_{cb}}=A \lambda,
\tilde{V_{c{b}\prime}}=q\lambda^{2}e^{-\iota\delta_{c{b}\prime}},\\
\tilde{V_{td}}=A \lambda^{3}(1- C e^{\iota \delta_{ub}})+r\lambda^{4}(q e^{-\iota\delta_{c{b}\prime}}-p e^{-\iota\delta_{u{b}\prime}}),\\
\tilde{V_{ts}}=-A\lambda^{2}-q r\lambda^{3}e^{-\iota\delta_{c{b}\prime}} +\frac{A}{2}\lambda^{4}(1+r^{2}C e^{\iota \delta_{ub}}),
\tilde{V_{tb}}=1- \frac{r^{2} \lambda^{2}}{2},\\
\tilde{V_{t{b}\prime}}=r\lambda,
\tilde{V_{{t}\prime d}}=\lambda^{3}(q e_{\iota\delta_{c{b}\prime}})+ Ar\lambda^{4}(1+Ce^{\iota \delta_{ub}}),\\
\tilde{V_{{t}\prime s}}=q\lambda^{2}e^{-\iota\delta_{u{b}\prime}}+ Ar\lambda^{3}+\lambda^{4}(-p e^{-\iota\delta_{u{b}\prime}}+\frac{q}{2}e_{\iota\delta_{c{b}\prime}}+ \frac{qr^{2}}{2} e_{\iota\delta_{c{b}\prime}} ),\\
\tilde{V_{{t}\prime b}}=-r\lambda \quad and \quad
\tilde{V_{{t}\prime{b}\prime}}=1- \frac{r^{2}\lambda^{2}}{2}$.
\newpage
\noindent where all the elements of $U_{CKM}$ are unitary up to ${\cal O}(\lambda^4)$. 

The above expansion corresponds to the Wolfenstein
parametrization with $C =\sqrt{\rho^{2} + \eta^{2}}$ and $\delta_{ub} = \tan^{-1}(\frac{\eta}{\rho})$. Constraints on all the elements of $CKM_{4}$ matrix formulated using DK parametrization can be obtained by using the unitarity
of the $CKM_4$ matrix. Through a variety of independent measurements, the SM $3\times 3$
submatrix have been found to be approximately unitary. The values of $CKM_{4}$ parameters are taken from \cite{Alok,24,25,26} where they perform the $\chi^{2}$-fit at two values of $t\prime$ mass i.e. $m_{t}\prime = 400 GeV$ $\&$ $m_{t}\prime = 600 GeV$. The $4^{th}$-generation quark masses
are constrained to a narrow band, which increases the predictability of the $SM_4$. Along with that, they have also generated a fit for the 4 Wolfenstein parameters of the CKM matrix in the SM,
in order to check for consistency with the standard fit. The results summarised are shown in table~{\bf\ref{my-label}}.
\begin{table}[hbtp]
\centering
\begin{tabular}{|c|c|c|c}
\hline 
 \textbf{$CKM_{4}$ Parameters} & \textbf{$m_{t}\prime=400 GeV$} & \textbf{$m_{t}\prime=600 GeV$}  \\ 
 \hline \hline
$\lambda$ &$ 0.227 \pm 0.001 $&$ 0.227 \pm 0.001   $ \\
\hline 
A &$ 0.801 \pm 0.022 $&$ 0.801 \pm 0.002  $ \\
 \hline
C &$ 0.38 \pm 0.04$ & $0.42 \pm 0.04  $  \\ 
\hline
$\delta_{ub}$ & $1.24 \pm 0.23$ &$ 1.22 \pm 0.24 $\\
 \hline
p &$ 1.45 \pm 1.20$ & $1.35 \pm 1.53 $ \\
 \hline
q &$ 0.16 \pm 0.12 $& $0.12 \pm 0.07 $ \\
\hline
r &$ 0.30 \pm 0.37$ &$ 0.19 \pm 0.27 $\\
 \hline
$\delta_{ub\prime}$ & $1.21 \pm 1.59 $& $1.32 \pm 1.76 $ \\
 \hline
$\delta_{cb\prime}$ & $1.10 \pm 1.64$ & $1.25 \pm 1.81 $ \\
\hline 
\end{tabular}
\caption{Numerical values of all the parameters used in $U_{CKM_{4}}$ matrix \cite{Alok,24,25,26}.}
\label{my-label}
\end{table}
On the other hand, the lepton mixing matrix $U_{PMNS}$ for our analysis we have taken a basis where the charged lepton
mass matrix is diagonal. Therefore, the lepton mixing matrix is
simply $U_{PMNS} = U$. So any complex symmetric $4 \times 4$ light neutrino mass matrix can be written as
\begin{equation}
M_{\nu}= U M_{\nu}^{diag} U^{T},
\end{equation}
\noindent where $M_{\nu}^{diag}= diag(m_{1},m_{2},m_{3},m_{4}$) is the diagonal form
of the light neutrino mass matrix. The diagonalizing matrix U is the $4 \times 4$ version of the
PMNS leptonic mixing matrix which is parametrized as
{\small
\begin{equation}\nonumber
U= \tilde{R_{34}}R_{24}\tilde{R_{14}}R_{23}\tilde{R_{13}}R_{12},
\end{equation}
}
\noindent where the rotation matrices R, $\tilde R$ can be further parametrized as(for example $R_{24}$ and $\tilde R_{34}$) 
 \[
R_{24} =
\begin{bmatrix}
1 & 0 & 0&0\\
0 & 1 &0&0\\
0  &C_{24}&1&S_{24}\\
0 &-S_{24}& 0&C_{24}
\end{bmatrix}
,
\]
 \[
\tilde {R_{34}} =
\begin{bmatrix}
1 & 0& 0&0\\
0 & 1 &0 &1\\
0 & 0& C_{34}& S_{34}e^{\iota \phi}\\
0 & 0&-  S_{34}e^{\iota \phi}& C_{34}
\end{bmatrix}
,
\]
\noindent where $C_{ij}\equiv \cos\theta_{ij}$, $S_{ij}\equiv \sin\theta_{ij}$ and $\tilde S_{ij}\equiv S_{ij} e^{-\iota \phi_{ij}}$ and here,
$\phi_{ij}$ are the lepton Dirac CP phases. These phases are generalised as $\phi$ as these are unconstrained and we used the same range of spread for all [$0-2\pi$] with flat distribution.

However, the values of the $U_{PMNS_{4}}$ 
angles are taken as under at 1-$\sigma$ level \cite{king}
\begin{eqnarray}\label{pmnsa}
 \sin^{2}\theta_{13}=0.0218_{-0.0010}^{+0.0010},\\ \nonumber
 \sin^{2}\theta_{12}= 0.304_{-0.012}^{+0.013},\\ \nonumber
 \sin^{2}\theta_{23}=0.452_{-0.028}^{+0.052}.
%  \phi= (306^{\circ})_{-70}^{+39}.
\end{eqnarray}

The value of CP violation phases $\phi$ have been kept open varying freely between ($0- 2 \pi$) and the values used for $\theta_{14}$, $\theta_{24}$ and $\theta_{34}$ are assumed to vary freely between ($0- \pi/4$). The reason behind this specific limit ($0- \pi/4$) is that all the values obtained using our reference experiments i.e. No$\nu$A, MINOS, SuperK, and IceCube-DeepCore \cite{experi,30,31,32,33} vary between this similar range so instead of taking a specific value we have take that whole range in our model. The table {\bf\ref{label1}} below shows all the upper limits obtained from various experiments.
\vspace{0.2cm}
\begin{table}[!hbtp]
\centering
\begin{tabular}{|c|c|c|c|c|c|}
\hline 
 \textbf{Experiment} & \textbf{$\theta_{24}^{PMNS_{4}}$} & \textbf{$\theta_{34}^{PMNS_{4}}$} & $\mid U_{\mu 4}\mid^{2}$ & $\mid {U_{\tau 4}}\mid^{2}$ \\ 
 \hline \hline
No$\nu$A & 20.8 & 31.2 &  0.126 & 0.268 \\
\hline 
MINOS & 7.3 & 26.6 & 0.016 & 0.20 \\
 \hline
SuperK & 11.7 & 25.1 & 0.041  & 0.18 \\ 
\hline
IceCube-DeepCore & 19.4 & 22.8 & 0.11 & 0.15\\
 \hline
\end{tabular}
\caption{{\textit The upper limits obtained from NO$\nu$A, MINOS, Super-Kamiokande, IceCube and IceCube-
DeepCore.}}
\label{label1}
\end{table}

 After performing the Monte Carlo simulations we estimated the texture of the correlation matrix ($V_{c_{4}}$) for two different values of $m_{{t}\prime}= $ $400 GeV$ $\&$ $600 GeV$ (where $m_{{t}\prime}$ is the mass of ${t}\prime$) and implemented the same matrix in our inverse equation and obtained the constrained results for the sterile neutrino parameters.
 We obtained predictions for  $\theta_{24}$ and $\theta_{34}$ and then compared our results with the current experimental bounds given by No$\nu$A, MINOS, SuperK, and IceCube-DeepCore \cite{experi,30,31,32,33} experiments.

\section{Results } \label{sec:4}
We have divided our results in two parts i.e. for $m_{{t}\prime}= 400 GeV$ and  $m_{{t}\prime}= 600 GeV$. The $PMNS_{4}$ matrix obtained in case of $m_{{t}\prime} = 400 GeV$ is

\[
U_{PMNS_{4}}=
  \begin{bmatrix}
 0.5596... 0.5625 & 0.2235... 0.2314 & 0.1520... 0.1770 & 0.0131... 0.1975 \\
 0.3339... 0.3370&0.4181... 0.3370&0.4181... 0.4224&0.0180... 0.1732 \\
 0.0375... 0.1394&0.3310... 0.3485&0.4379... 0.4594&0.0076... 0.4507 \\
 0.0076... 0.01541&0.0062... 0.2154&0.0050... 0.2832&0.7332... 0.7596 
\end{bmatrix}
,\]

\noindent and the best fit obtained using a normal distribution of the observables including both quark and lepton parameters is in the form of the matrix below

\[
U_{PMNS_{4}}=
  \begin{bmatrix}
 0.5614 & 0.2281 & 0.1642 & 0.1016 \\
 0.5450 & 0.3353 & 0.4203 & 0.0970 \\
 0.1006 & 0.3447 & 0.4557 & 0.1997 \\
 0.0775 & 0.1118 & 0.1406 & 0.7571 
\end{bmatrix}
.\]

 In case  of $m_{{t}\prime} = 600 GeV$ $PMNS_{4}$ is

 \[
U_{PMNS_{4}}=
  \begin{bmatrix}
 0.5596... 0.5630 & 0.2259... 0.2337 & 0.1538... 0.1725 & 0.0140... 0.2240 \\
 0.5422... 0.5449 & 0.3304... 0.3325 & 0.4176... 0.4204 & 0.0130... 0.1538 \\
 0.07385... 0.1243 & 0.3345... 0.3450 & 0.4475... 0.4562 & 0.0070... 0.3784 \\
 0.0118... 0.1776& 0.0042... 0.1919& 0.0049... 0.2299 & 0.7438... 0.7565
  \end{bmatrix}
,\]

\noindent and the best fit obtained is
 \[
U_{PMNS_{4}}=
  \begin{bmatrix}
0.5619 & 0.2307 & 0.1631 & 0.1052 \\
 0.5437 & 0.3314 & 0.4190 & 0.0932 \\
 0.0994 & 0.3424 & 0.4548 & 0.1637 \\
 0.0783 & 0.1074 & 0.1044 & 0.7553
  \end{bmatrix}
.\]

As per our model procedure, in order to constrain the sterile neutrino parameters i.e. $\theta_{24}^{PMNS}$,  $\theta_{34}^{PMNS}$ $\mid U_{\mu 4} \mid^{2}$ and $\mid U_{\tau 4} \mid^{2}$ we have
used the inverse equation {\bf\ref{eq:pmnsinv}}.

\subsection{Predictions for \texorpdfstring{$\theta_{24}^{PMNS}$}{\theta\_{24}\^{PMNS}} and \texorpdfstring{$\theta_{34}^{PMNS}$}{\theta\_{34}\^{PMNS}}}\label{sec:pmns}

We have investigated the implication of the 
non-trivial structure of the $V_{c_{4}}$ correlation matrix in the light of the latest results of various experiments. After using the $CKM_{4}$ and $PMNS_{4}$ parametrization we obtain the structure of $V_{c_{4}}$ from equation {\bf\ref{2}}, the analytical equations used in order to calculate the values of sterile neutrino parameters analytically as well as numerically are as follows
\begin{equation}\label{sterile}
\begin{aligned}
\theta_{24}=  \cos (\text{$\theta_{14} $}) \sin (\text{$\theta_{24} $}) \\ 
\theta_{34}= e^{-\iota \phi } \cos (\text{$\theta _{14}$}) \cos (\text{$\theta_{24} $}) \sin
   (\text{$\theta _{34}$}) \\ 
\mid U_{\mu 4}\mid^2= \left(1-\sin ^2(\text{$\theta_{14} $})\right) \sin ^2(\text{$\theta_{24}
   $})\\ 
  \mid U_{\tau 4}\mid^2= \left(1-\sin ^2(\text{$\theta_{14} $})\right) \left(1-\sin
   ^2(\text{$\theta_{24} $})\right) \sin ^2(\text{$\theta_{34} $})
   \end{aligned}
\end{equation}

The tables {\bf\ref{label2}} and {\bf\ref{label}} below show the comparison of upper limits obtained above with the four different experimental results.

\begin{table}[!hbtp]
  \centering

\begin{tabular}{|c|c|c|c|c|c|}

\hline 
 \textbf{Parameters} & \textbf{$\theta_{24}^{PMNS_{4}}$} & \textbf{$\theta_{34}^{PMNS_{4}}$}  \\ 
 \hline \hline
For $m_{{t}\prime} = 400 GeV$  & $6.57^{\circ}-23.36^{\circ}$ & $1.53^{\circ}-31.59^{\circ}$  \\
\hline 
For $m_{{t}\prime} = 600 GeV$ & $6.87^{\circ}-23.15^{\circ}$ & $3.78^{\circ}-32.40^{\circ}$ \\
 \hline
 \textbf{Parameters} &  $\mid U_{\mu 4}\mid^{2}$ & $\mid {U_{\tau 4}}\mid^{2}$\\
 \hline
 For $m_{{t}\prime} = 400 GeV$  &$0.0003-0.0300$ & $ 0.00-0.2031$\\
  \hline
 For $m_{{t}\prime} = 600 GeV$  & $ 0.0001-0.0236$  & $0.00-0.1432$\\
  \hline
\end{tabular}
\caption{ The limits obtained on sterile mixing angles.}
\label{label2}

\end{table}

\begin{table}[!hbtp]
  \centering

\begin{tabular}{|c|c|c|c|c|c|}

\hline 
 \textbf{Experiment} & \textbf{$\theta_{24}^{PMNS_{4}}$} & \textbf{$\theta_{34}^{PMNS_{4}}$} & $\mid U_{\mu 4}\mid^{2}$ & $\mid {U_{\tau 4}}\mid^{2}$ \\ 
 \hline \hline
No$\nu$A & 20.8 & 31.2 & 0.126 & 0.268   \\
\hline 
MINOS & 7.3 & 26.6 &0.016 & 0.20  \\
 \hline
SuperK & 11.7 & 25.1 &0.041 & 0.18   \\ 
\hline
IceCube-DeepCore & 19.4 & 22.8 & 0.11 & 0.15\\
 \hline
  \textbf{QLC Model} & \textbf{$\theta_{24}^{PMNS_{4}}$} & \textbf{$\theta_{34}^{PMNS_{4}}$} & $\mid U_{\mu 4}\mid^{2}$ & $\mid {U_{\tau 4}}\mid^{2}$ \\ 
 \hline
QLC(400 GeV) &23.36  & 31.59 &0.030 & 0.203 \\
 \hline
QLC(600 GeV) & 23.15 & 32.40 & 0.024 & 0.143\\
\hline

\end{tabular}
\caption{ The upper limits of sterile mixing
parameters obtained from model(QLC) and from NO$\nu$A, MINOS, Super-Kamiokande and IceCube-
DeepCore.}
\label{label}
\end{table}

\begin{figure*}[!hbt]
\minipage[t]{0.45\linewidth}
 
\includegraphics[width=\textwidth]{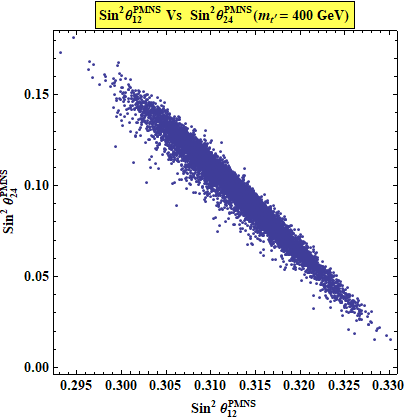}
\endminipage\hfill
\vspace{0.08cm}
\minipage[t]{0.45\linewidth}
 
\includegraphics[width=\textwidth]{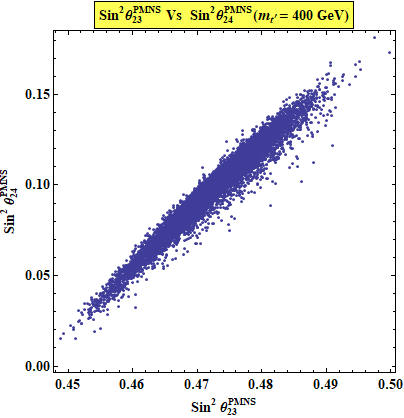}
 \endminipage\hfill
\caption{ Correlation plot for $\sin^{2}\theta_{24}^{PMNS}$ against $\sin^{2}\theta_{12}^{PMNS}$(Left) and $\sin^{2}\theta_{23}^{PMNS}$(Right) for $m_{{t}\prime}= 400 GeV$ .}
 \label{5o1}
\end{figure*}

\begin{figure*}[!hbt]
 \minipage[t]{0.45\linewidth}
 
\includegraphics[width=\textwidth]{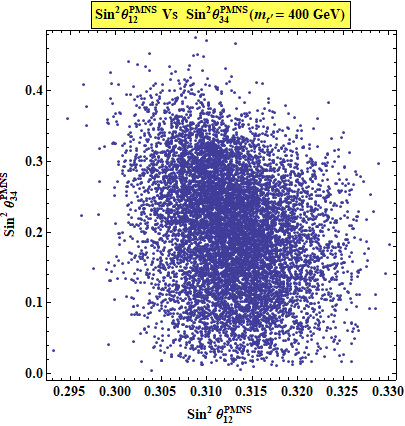}
\endminipage\hfill
\vspace{0.08cm}
\minipage[t]{0.45\linewidth}
 
\includegraphics[width=\textwidth]{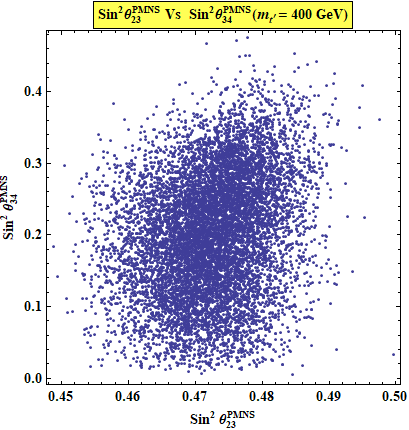}
 \endminipage\hfill
\caption{ Correlation plot for $\sin^{2}\theta_{34}^{PMNS}$ against $\sin^{2}\theta_{12}^{PMNS}$(Left) and $\sin^{2}\theta_{23}^{PMNS}$(Right) for $m_{{t}\prime}= 400 GeV$ .}
 \label{5o2}
\end{figure*}

\begin{figure*}[!hbt]

\minipage[t]{0.45\linewidth}
 
\includegraphics[width=\textwidth]{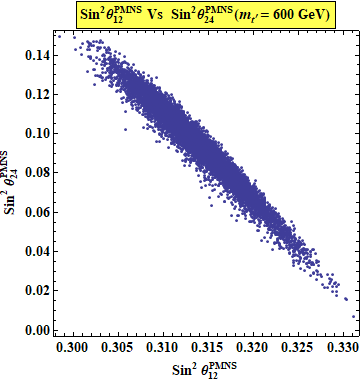}
\endminipage\hfill
\vspace{0.08cm}
\minipage[t]{0.45\linewidth}
 
\includegraphics[width=\textwidth]{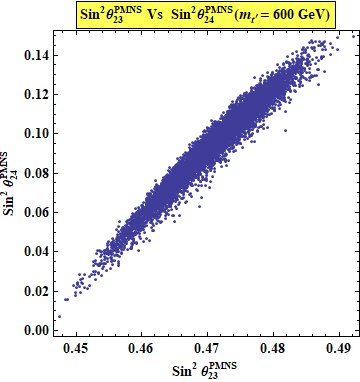}
 \endminipage\hfill
 
\caption{ Correlation plot for $\sin^{2}\theta_{34}^{PMNS}$ against $\sin^{2}\theta_{12}^{PMNS}$(Left) and $\sin^{2}\theta_{23}^{PMNS}$(Right) for $m_{{t}\prime}= 600 GeV$ .}
 \label{5o3}
\end{figure*}

\begin{figure*}[!hbt]
\minipage[t]{0.45\linewidth}
 
\includegraphics[width=\textwidth]{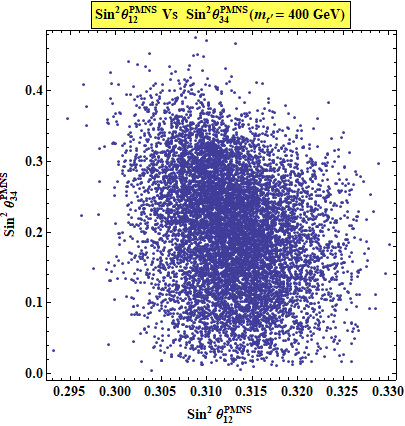}
\endminipage\hfill
\vspace{0.8cm}
\minipage[t]{0.45\linewidth}
 
\includegraphics[width=\textwidth]{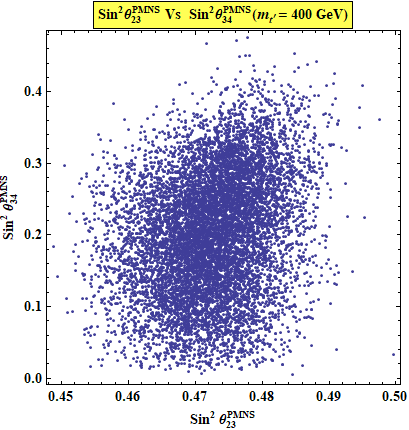}
 \endminipage\hfill
\caption{ Correlation plot for $\sin^{2}\theta_{34}^{PMNS}$ against $\sin^{2}\theta_{12}^{PMNS}$(Left) and $\sin^{2}\theta_{23}^{PMNS}$(Right) for $m_{{t}\prime}= 600 GeV$ .}
 \label{5o4}
\end{figure*}

During the analysis of the above results, one can observe that values obtained by QLC(400/600 GeV) lies close to the results obtained from the No$\nu$A and IceCube-DeepCore experiments. Numerically in all four equations {\bf\ref{sterile}}, the effect of the sterile neutrino mixing parameters can be clearly seen on one another. We report interrelated behaviour between $\sin^{2}\theta_{34}^{PMNS}$ and $\sin^{2}\theta_{24}^{PMNS}$ with the help of $3-\sigma$ scattered plots in space of two sterile neutrino mixing angles $\theta_{34}^{PMNS}$ and $\theta_{24}^{PMNS}$ with the varying range of $\sin^{2}\theta_{12}^{PMNS}$ and $\sin^{2}\theta_{23}^{PMNS}$. The correlation between the solar and atmospheric mixing parameters ($\sin^{2}\theta_{12}^{PMNS}$ and $\sin^{2}\theta_{23}^{PMNS}$) and the active-sterile mixing parameter ($\sin^{2}\theta_{24}^{PMNS}$ and $\sin^{2}\theta_{24}^{PMNS}$) is shown in figures {\bf\ref{5o1}}, {\bf\ref{5o2}}, {\bf\ref{5o3}} and {\bf\ref{5o4}}. Here, we note that the small mixing with sterile neutrinos will in general modify mixing scenarios. The mixing angles of active and sterile neutrinos are of order $ e/m_{s}$, where $\lq e \rq$ is any of the entries ${(U^{4 \times 4}_{\nu} )}_{f_s}$ with $f = e, \mu, \tau$ and $\lq m_s \rq$ is the sterile neutrino mass. Deviations from initial mixing angles $\theta_{12}^{PMNS}$,$\theta_{13}^{PMNS}$ and $\theta_{23}^{PMNS}$ are of the same order.

\begin{figure*}[!hbt] 
\minipage[t]{0.45\linewidth}
 
\includegraphics[width=\textwidth]{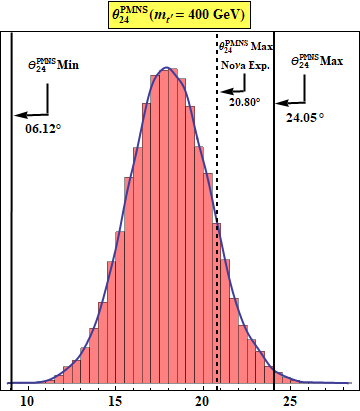}
\endminipage\hfill
\vspace{0.08cm}
\minipage[t]{0.45\linewidth}
 
\includegraphics[width=\textwidth]{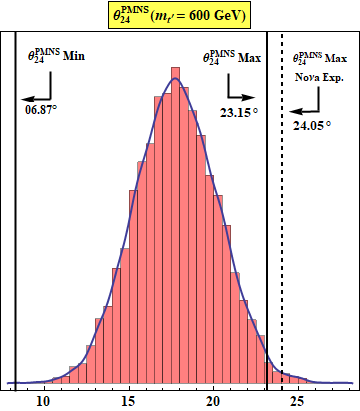}
 \endminipage\hfill
\caption{ Probability density distribution of $\theta_{24}^{PMNS}$ for 
$m_{{t}\prime}= 400 GeV$(Left) and $m_{{t}\prime}= 600 GeV$(Right) }
 \label{24theta}
\end{figure*}

\begin{figure*}[!hbt] 
\minipage[t]{0.45\linewidth}
 
\includegraphics[width=\textwidth]{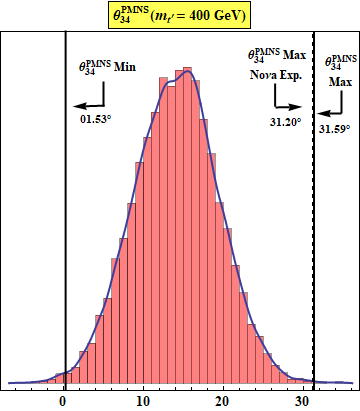}
\endminipage\hfill
\vspace{0.08cm}
\minipage[t]{0.45\linewidth}
 
\includegraphics[width=\textwidth]{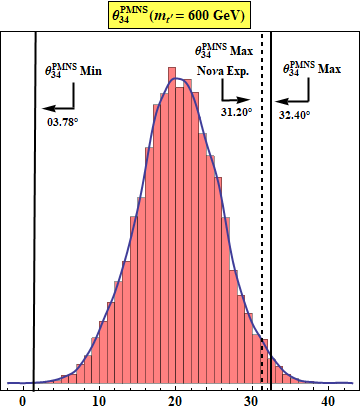}
 \endminipage\hfill
\caption{ Probability density distribution of $\theta_{34}^{PMNS}$ for 
$m_{{t}\prime}= 400 GeV$(Left) and $m_{{t}\prime}= 600 GeV$(Right) }
 \label{34theta}
\end{figure*}

As our results vary upon the value $m_{{t}\prime}= 400 GeV$ $\&$ $600 GeV$
we have obtained histograms of the probability density 
function for  $\theta_{24}^{PMNS}$ and $\theta_{34}^{PMNS}$ for both $400 GeV$ $\&$ $600 GeV$, respectively and show comparison between the two. In the figure {\bf\ref{24theta}}and {\bf\ref{34theta}}
we have shown this quite nicely the left panel of the figure is for
$\theta_{24}^{PMNS}$ $\&$ $\theta_{34}^{PMNS}$ for $m_{{t}\prime}= 400 GeV$, whereas the
right panel shows the $\theta_{24}^{PMNS}$ $\&$ $\theta_{34}^{PMNS}$ for $m_{{t}\prime}= 600 GeV$ respectively.
We have analysed numerically the normal distributions of both the mixing angles through histograms in the figure {\bf\ref{24theta}} and {\bf\ref{34theta}}. We have compared our results for the upper values from No$\nu$A experiment. Here the dashed lines are for $\theta_{24}^{PMNS}$ $\&$ $\theta_{34}^{PMNS}$ (No$\nu$A experiment), the thick solid lines are the 3-$\sigma$ upper and lower values of $\theta_{24}^{PMNS}$ and $\theta_{34}^{PMNS}$ obtained using the QLC model.

\begin{figure*}[!hbt]
\minipage[t]{0.45\linewidth}
 
\includegraphics[width=\textwidth]{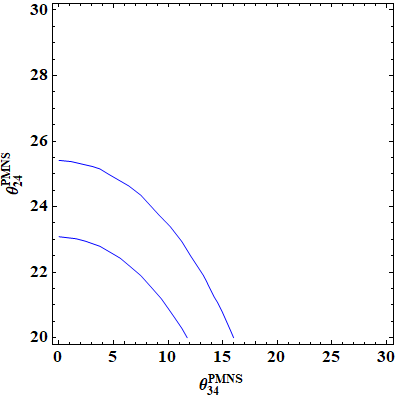}
\endminipage\hfill
\vspace{0.08cm}
\minipage[t]{0.45\linewidth}
 
\includegraphics[width=\textwidth]{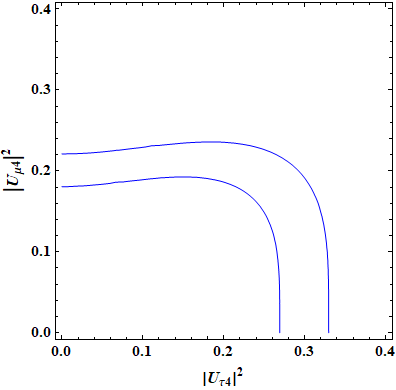}
 \endminipage\hfill
\caption{ Contour plots between $\theta_{24}^{PMNS}$ and $\theta_{34}^{PMNS}$ (Left) $\&$ $\mid {U_{\mu 4}}\mid^{2}$ and $\mid {U_{\tau 4}}\mid^{2}$ (Right) where $\theta_{14}^{PMNS}$ is assumed to vary between $0- \pi/4$.}
 \label{5o}
\end{figure*}

In order to analyse their impact more accurately we have made contour plots in the figure {\bf\ref{5o}}. If, the mixing angles are expressed in terms of the relevant matrix elements(eq. {\bf\ref{sterile}}) then the limits on $\theta_{34}$ and $\theta_{24}$ becomes $\mid {U_{\mu 4}}\mid^{2}$ and $\mid {U_{\tau 4}}\mid^{2}$ at the $99.7\%$ C.L. i.e. $3\sigma$ range. This whole analysis is very less sensitive to $\theta_{14}$ which is constrained to be small by reactor experiments \cite{34} as well as the QLC model analysis via which the value obtained is smaller as compared to the other two angles. 
The contour plots shown in the corresponding figures {\bf\ref{5o}} are depicting the correlation between $\theta_{34}$ and $\theta_{24}$ for $m_{{t}\prime}= $ $400 GeV$ $\&$ $600 GeV$  and $\mid {U_{\mu 4}}\mid^{2}$ and $\mid {U_{\tau 4}}\mid^{2}$ again for $m_{{t}\prime}=$ $400 GeV$ $\&$ $600 GeV$, respectively. These plots clearly depict constrained ranges of  $\theta_{34}$ and $\theta_{24}$ as well as $\mid {U_{\mu 4}}\mid^{2}$ and $\mid {U_{\tau 4}}\mid^{2}$ which are comparable to the experimental results obtained by No$\nu$A and IceCube-DeepCore \cite{19,ic}.

\section{ Conclusions} \label{sec:5}
In the desire to understand the depth of the quark and lepton
world and their varying scenario, the quark-lepton symmetry and unification field have drawn a lot of attention of many researchers in recent years. Out of the different aspects that
imply the symmetry and unification in the quark and lepton
sectors, the QLC relations between the mixing angles of
the $U_{CKM}$ and $U_{PMNS}$ matrices have been considered very
interesting and suggestive. Motivated by previous works towards its understanding, in this paper, we have made an attempt to take forward our previous work in a completely new direction to explain the QLC Model with sterile neutrinos. 
We have introduced a totally new approach of QLC involving the existence of sterile neutrino using 3+1 scenario. 

The detailed analysis is done for the non-trivial relation between the $U_{PMNS_{4}}$ and $U_{CKM_{4}}$ mixing matrices along with the phase mismatch between the quarks and leptons via $\psi_{4}$ the diagonal matrix phase. Using all the parameters that are available from the global data analysis we have investigated the structure of the correlation matrix $V_{c_{4}}$, numerically for two different values of $m_{{t}\prime}=$ $400 GeV$ $\&$ $600 GeV$. We have obtained results for two sterile neutrino mixing angles i.e.
$\theta_{24}^{PMNS}$ and $\theta_{34}^{PMNS}$ and two elements of the $PMNS_{4}$ mixing matrix i.e. $\mid {U_{\mu 4}}\mid^{2}$ and $\mid {U_{\tau 4}}\mid^{2}$ then compared the upper limits with different experiments mentioned above. We have compared just the upper limits because only the upper limits of all the parameters are available collectively in the recent results from these experiments \cite{experi,30,31,32,33}. The whole analysis gives us the hint about the relevance of the sterile neutrinos with the quark-lepton unification and the model we have been using i.e. Quark-Lepton Complementarity. 
The complete understanding of such wide dissimilarity between the quark and lepton
mixing patterns are considered to be one of the biggest challenges for the physics beyond the
standard model.

In present world keeping in view all experiments these results are still favoured by
the present experimental data within their measurement errors and if these QLC relations are not accidental, they strongly suggest the common connection between quarks and leptons at some high energy scale.
Although it is very hard to understand these type of relations in ordinary bottom-up
approaches, where the quarks and leptons are treated separately with no specific connection is seen
between them. We do require some top-down approaches like the Grand Unified Theories(GUT)
which sometimes also unify quarks and leptons and provide a framework to construct a model
in which QLC relation can be embedded in a natural way. With this endeavor to perform numerical simulations in order to investigate the sterile neutrino mixing angles, one might have an eye to look forward towards the understanding of the QLC model in a better way. The connoisseur eye and deep knowledge with some theoretical ground can explain QLC relations precisely. The results obtained numerically and analytically in this paper can be noticed in a good agreement with the experimental data.
\vspace{0.5cm}

\noindent {\large {\bf Acknowledgements}}

We thank Dr. Surender Verma for his valuable comments and suggestions in the completion of the work. B.C. Chauhan acknowledges the financial support provided by the University Grants Commission(UGC), Government of India vide Grant No. UGC MRP-MAJOR-PHYS-2013-12281. We thank IUCAA for providing research facilities during the completion of this work.

\end{document}